# Strain-induced quantum ring hole states in a gated vertical quantum dot

Jun Liu[1], A. Zaslavsky[2], L. B. Freund[2]

[1]*Department of Physics, Brown University, Providence, RI 02912, USA*

[2]*Division of Engineering, Brown University, Providence, RI 02912, USA*

We have experimentally investigated the hole states in a gated vertical strained Si/SiGe quantum dot. We demonstrate the inhomogeneous strain relaxation on the lateral surface creates a ring-like potential near the perimeter of the dot, which can confine hole states exhibiting quantum ring characteristics. The magnetotunneling spectroscopy exhibits the predicted periodicity of energy states in $\phi/\phi_0$, but the magnitude of the energy shifts is larger than predicted by simple ring theory. Our results suggest a new way to fabricate and study quantum ring structures.

Quantum ring structures exhibit many interesting physical phenomena. The closed loop geometry of quantum rings [1] gives rise to energy spectra periodic in the enclosed magnetic flux $\phi$ with period $\phi_0 = h/e$, in analogy to Aharonov-Bohm effect [2]. Various experiments [3-5] have investigated properties of quantum rings. Most recently, both optical spectroscopy [6,7] and transport spectroscopy [8] have been employed to study energy spectra of electrons in rings fabricated in III-V semiconductors. However, so far there has been no experiment investigating hole states in quantum rings, while the properties of quantum ring hole states might differ because of the anisotropy and nonparabolicity in hole dispersions. Here we report the first unambiguous observation of quantum ring hole states, which are confined by inhomogeneous-strain-induced ring-like potential to the perimeter of a vertical $p$-Si/Si$_{1-x}$Ge$_x$ quantum dot. The magnetotunneling spectroscopy demonstrates the predicted periodicity of energy states in the enclosed magnetic flux $\phi$ with period $\phi_0$, but the magnitude of the energy shifts is larger than predicted by simple ring theory [9]. Our results also suggest a new way to fabricate and study quantum ring structures.

Lattice mismatch between Si and SiGe introduces compressive biaxial strain in the SiGe epitaxial layer, which remains homogeneous as long as the SiGe layer thickness stays below the critical thickness $h_c$ [10]. This strain affects the band structure, lifting the degeneracy of the heavy-hole (HH) and light-hole (LH) valence band-edges and decreasing the bandgap [11]. Consequently inhomogeneous strain induces a nonuniform quantum-mechanical potential for carriers [12-14]. Finite element simulations [15] predict that when a vertical quantum dot is etched out of strained $p$-Si/SiGe heterostructure, outward lattice displacement will relax the strain near the sidewall [15-17], leading to radially inhomogeneous strain, and hence an inhomogeneous potential for holes [18, 19]. Surprisingly, the simulations predict that as the size of the vertical quantum dot is reduced to the deep submicron region, there is a ring-like confining potential for holes near the perimeter of the quantum dot, which can lead to quantum ring hole states [19, 20]. However, although the high magnetic field measurements of the vertical quantum dot [20] provided evidence for the existence of ring-like confinement consistent with the finite element simulations, the most fundamental characteristic of quantum ring structures and the conclusive proof of their existence, namely, the energy spectrum periodic in the enclosed magnetic flux, was not reported yet.

To unambiguously identify and study the inhomogeneous-strain-induced quantum ring hole states, we fabricated a vertically gated quantum dot shown schematically in Fig. 1(a). Gate voltage can be applied to reduce the effective size of the quantum dot and reveal the inhomogeneous-strain-induced confinement by depletion from the side. Unlike GaAs-based gated vertical quantum dots [21, 22], where a large depletion exists due to Fermi-Level pinning at the surface, our device is oxide-insulated [23], with a much smaller surface density of states which does not cause significant depletion at the sidewall [19].

The submicron vertically gated quantum dot was fabricated from high quality strained *p*-Si/SiGe triple-barrier resonant tunneling material [24]. Figure 1(a) shows the schematic of the cross section of the device. Electron-beam lithography and reactive ion etching were used to fabricate a submicron pillar of nominal radius $r$ = 160 nm. A relatively thick gate oxide of ~200 nm in thickness was then deposited conformally, followed by annealing in $N_2$ to reduce the interface states. Self-aligned Al gate of 60 nm in thickness was deposited around the double quantum well region of the quantum dot, followed by another thick layer of $SiO_2$ for isolation. A bias between the top contact and the substrate aligns occupied hole states in the emitter electrode with the quantized hole states in the Si/SiGe quantum wells. Figure 1(b) shows the heavy-hole (HH) and light-hole (LH) branches in the two quantum wells couple due to the thin middle barrier. Therefore the *I(V)* characteristics contain four features corresponding to tunneling to the coupled subbands [24]. On top of these features there is a fine structure corresponding to tunneling through the states confined by the inhomogeneous-strain-induced lateral potential in the wells [20].

When magnetotunneling measurements were performed at $T$ = 1.6 K in a weak magnetic field *B* parallel to the tunneling direction, we observed a periodic oscillation of a set of conductance peaks with a period of ~60 mT as shown in Fig. 2, which is equivalent to approximately one magnetic quantum flux enclosed by the quantum dot. We attribute these conductance peaks to tunneling into the quantum ring hole subbands confined by the in-plane inhomogeneous-strain-induced ring-like potential near the edge of the quantum well [20], shown in Fig. 3(a) as predicted by finite element analysis [15]. Simple one-dimensional ring theory predicts that in an infinitely thin quantum ring of radius *r* enclosing magnetic flux $\phi = \pi r^2 B$, the energy spectrum is given by [1]

$$E_l = \frac{\hbar^2}{2m^* r^2}(l + \frac{\phi}{\phi_0})^2 \qquad (1)$$

where $m^*$ is the effective mass, $\phi_0 = h/e$ is the magnetic flux quantum and $l$ is the angular momentum quantum number. Consequently the ground and excited state energies oscillate as a function of the magnetic flux with a period of $\phi_0$ [8]. Realistic quantum rings have finite width, which are able to accommodate multiple channels of states [25], each of which exhibits a similar energy spectrum. Accordingly, the density of states of realistic rings will also oscillate as a function of magnetic field. Since tunneling conductance directly measures the density of states available for tunneling in the double quantum wells, the measured oscillation period of ~60 mT in Fig. 2(b) translates into an effective radius of ~150 nm for the quantum ring, in excellent agreement with the nominal size of the quantum dot $r \sim 160$ nm. The $\phi/\phi_0$ periodicity and the large effective radius unambiguously exclude tunneling through impurity states or other extraneous effects.

Figure 3(a) presents our finite element simulation of the in-plane inhomogeneous-strain-induced potential for holes in the quantum wells [15]. Both the edge ring-like potential and the roughly parabolic central potential can produce quantized hole states available for tunneling, but only the quantum ring hole states are expected to respond to very weak $B$. This phenomenon is observed in the magnetotunneling conductance characteristics, in which except the quantum ring tunneling peaks, most conductance peaks do not shift in small magnetic fields. Figure 3(a) also shows the lowest energy levels of the laterally confined 1D ring subbands arising from the third hole subband of the coupled quantum wells, since the $B$-periodic conductance peaks are observed in the bias range corresponding to tunneling through the third subband, which has the lowest in-plane effective mass $m^* \sim 0.15$ $m_e$ [26]. Hole states confined in the parabolic central potential only introduce a linearly increasing contribution to the density of states in the quantum well; furthermore, finite element simulations indicate that only the edge quantum ring potential survives for sufficiently small quantum dots [15]. In order to clearly illustrate the contribution from the quantum ring states, we ignored the contribution from the centrally confined states in our numerical calculation of the density of states, using a simplified truncated parabolic potential (see inset in Fig. 3(b)) and single in-plane effective mass approximation $m^* \sim 0.15$ $m_e$ [26]. Several quantum ring subbands are shown confined by the truncated parabolic potential. Figure 3(b) shows the resulting calculated density of ring states. The peak positions of the density of states exhibit oscillations in magnetic field with a period of 60 mT, in excellent agreement with our experimental data. However, this preliminary analysis using simple quantum ring theory with $m^* \sim 0.15$ $m_e$ predicts an oscillation amplitude of order $\hbar^2/2m^*r^2 \sim 10^{-2}$ meV for ground state energy and only slightly larger amplitude for the overall density of ring states as shown in Fig. 3(b), while the oscillations in peak positions observed in our experiment correspond to ~0.3 - 1.1 meV in energy, which are converted from the bias using a self-consistent calculation analogous to Fig. 1(b). The very complicated nature of hole dispersions, which arises from the quantum well confinement and the interaction between the double quantum wells, may be the cause for the large oscillations we observed in experiment. Furthermore, we only considered single-particle levels in our model, ignoring many-body effects and possible scattering in the quantum ring, which could influence its energy spectrum [9]. In any case, a more thorough theoretical analysis of quantum ring hole states in the magnetic field incorporating the anisotropy and nonparabolicity of valence band remains to be developed.

We also investigated the gate effect on the quantum ring hole states. The side gate can be used to raise the energy of the edge quantum ring hole states or even pinch off these states completely, thus enables us to probe the spatial location of the quantum ring hole states. Figure 4(a) presents the $I(V, V_g)$ tunneling characteristics of the same device. As the gate voltage $V_g$ is increased from 0 to 20 V, the largest peak, which is corresponding to resonant tunneling into the fourth hole subband in the double quantum wells, is reduced by about 9 percent. This current reduction is equivalent to ~80 Å reduction in the effective radius of the vertical quantum dot. It is unsurprising to have such relatively small size reduction considering the thick gate oxide (~200 nm), the highly doped emitter and collector regions, and the relatively thin undoped active region [23]. Numerical calculations indicate that a 20 V gate-voltage can only cause approximately 50 Å depletion in the highly doped emitter. The depletion in the undoped active region is expected only slightly larger than in the emitter as shown in the inset of Fig. 4(a), due to the adjacent large acceptor density. Despite the small size reduction, the quantum ring states at the perimeter should be sensitive to $V_g$. Figure 4(b) shows the four conductance peaks corresponding to tunneling into the ring subbands shift to higher bias with higher $V_g$. On the other hand, the largest current peak at $V = 0.42$ V, which is dominated by tunneling through the core of the vertical quantum dot, remains unaffected. This observation indicates that the quantum ring states are indeed located near the perimeter of the quantum dot as the finite element simulations predict. Due to the small penetration length of the gate electric field, higher-lying quantum ring subbands with larger probability density near the perimeter are expected to be affected in a larger extent by $V_g$, again in agreement with our data.

In conclusion, we report the first unambiguous observation of quantum ring hole states, which are confined by inhomogeneous-strain-induced ring-like potential to the perimeter of an inhomogeneously strained SiGe quantum dot. The magnetotunneling spectroscopy exhibits the periodicity of energy states in the enclosed magnetic flux $\phi$ with period $\phi_0$, but the oscillation amplitude is larger than predicted by simple ring theory. The discrepancy implies the necessity in including the anisotropy and nonparabolicity of hole dispersions in quantum ring theories for holes in order to quantitatively interpret the experimental data. Our results also suggest the inhomogeneous strain relaxation in strained quantum structures can be utilized to fabricate and study quantum rings.

We thank H. Maris and B. Marston for helpful discussions. The work is supported by an NSF Career Award (DMR-9702725) and the NSF MRSEC Center (DMR-9632524). The fabrication facilities are also supported in part by Brown MRSEC.

Figure 1. (a) Schematic of cross section of the gated vertical quantum dot (not to scale). Hatched regions are metal, while the shaded region is silicon dioxide. Embedded in the diagram is the SEM picture of a small etched quantum dot for illustration. The bias $V$ is applied between the top contact pad and the $p^+$-Si substrate. Positive gate voltage $V_g$ is applied on the gate to squeeze the dot and reduce its size. (b) Self-consistently calculated valence band bending of the vertical quantum dot at a bias $V = 260$ mV, showing the alignment between the occupied emitter hole states and the coupled quantum well states. The four solid line are the coupled double quantum well bound states at $k_\parallel = 0$, calculated by six-band Luttinger-Kohn formalism, while the dotted lines show the HH and LH subbands that would exist in isolated SiGe quantum wells of the same width.

Figure 2. Magnetotunneling spectroscopy at $T = 1.6$ K. (a) Conductance through the ring states, indicated by arrows. Dotted lines are guides to the eye. The positions of the peaks corresponding to tunneling into the quantum ring hole subbands in the double quantum wells oscillate as a function of the magnetic field $B$; the rest of the conductance curve remain unchanged by small magnetic field ($B < 200$ mT). (b) The periodic oscillation of the positions of the conductance peaks with magnetic field, showing a period of ~60 mT. Dotted lines are guides to the eye.

Figure 3. Numerical simulations on the quantum ring hole states. (a) Finite element simulation of the strain relaxation and corresponding strain-induced confining potential in the quantum dot. The top curve shows the calculated radial strain component $\varepsilon_{rr}(r)$ on the mid-plane of the top 35 Å $Si_{0.8}Ge_{0.2}$ quantum well in the quantum dot (full biaxial strain corresponds to $\varepsilon_{rr} = 1$), while the bottom curve is the resulting in-plane hole confining potential as a function of $r$. The solid lines mark the lowest energy levels of the quantum ring subbands in the ring-like potential near the perimeter. (b) Density of states of the laterally quantized quantum ring hole states in the quantum dot. The numerical calculations use the simplified truncated parabolic ring potential shown in the left inset, with single effective mass approximation $m^* = 0.15$ $m_e$ calculated using the six-band Luttinger-Kohn model. The calculations assume ~0.2 meV Gaussian broadening of individual ring states, considering the finite temperature broadening at $T = 1.6$ K. The right inset plots the positions of the third peak (indicated by the arrow) of the calculated density of ring states with magnetic field.

Figure 4. *I(V, $V_g$)* characteristics. (a) Resonant tunneling *I(V, $V_g$)* of the vertically gated quantum dot at $T = 1.6$ K. The largest peak current is reduced by 9 percent at $V_g = 20$ V. The position of the largest peak is unaffected by the gate voltage $V_g$, because it is dominated by tunneling through the core of the quantum dot. The arrows indicate the bias range for observed ring state tunneling. The inset illustrates the equipotential (solid curve) in the active region when the gate voltage is applied (not to scale). (b) Conductance peaks through the ring states, showing the strong $V_g$-induced shifts of the ring hole subbands at the perimeter. Higher-lying subbands have larger shifts with $V_g$. Dotted lines are guides to the eye.

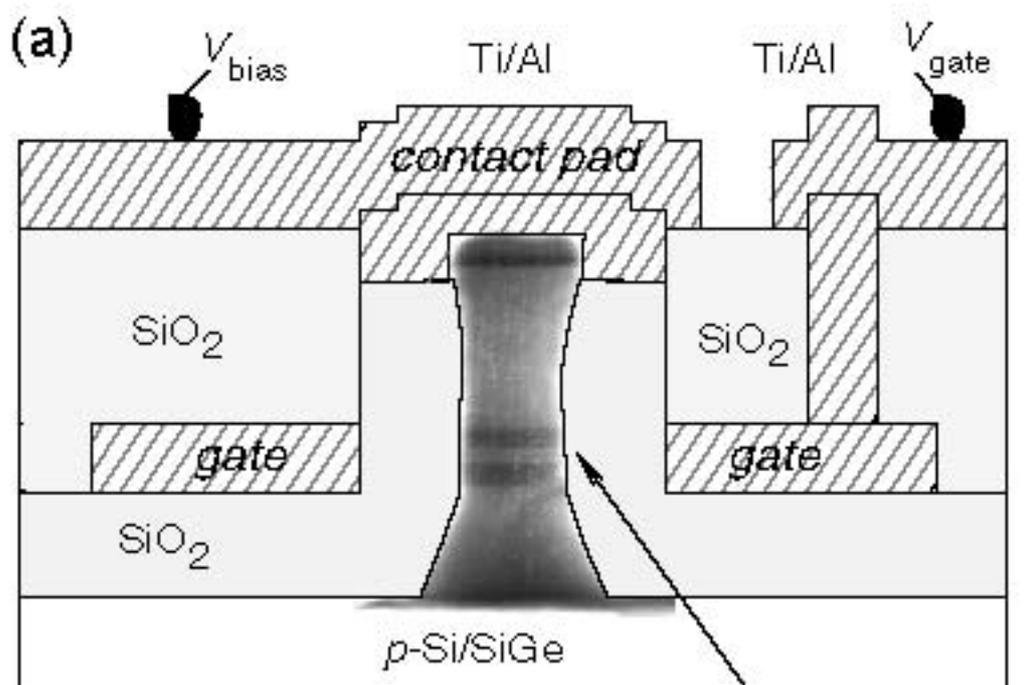
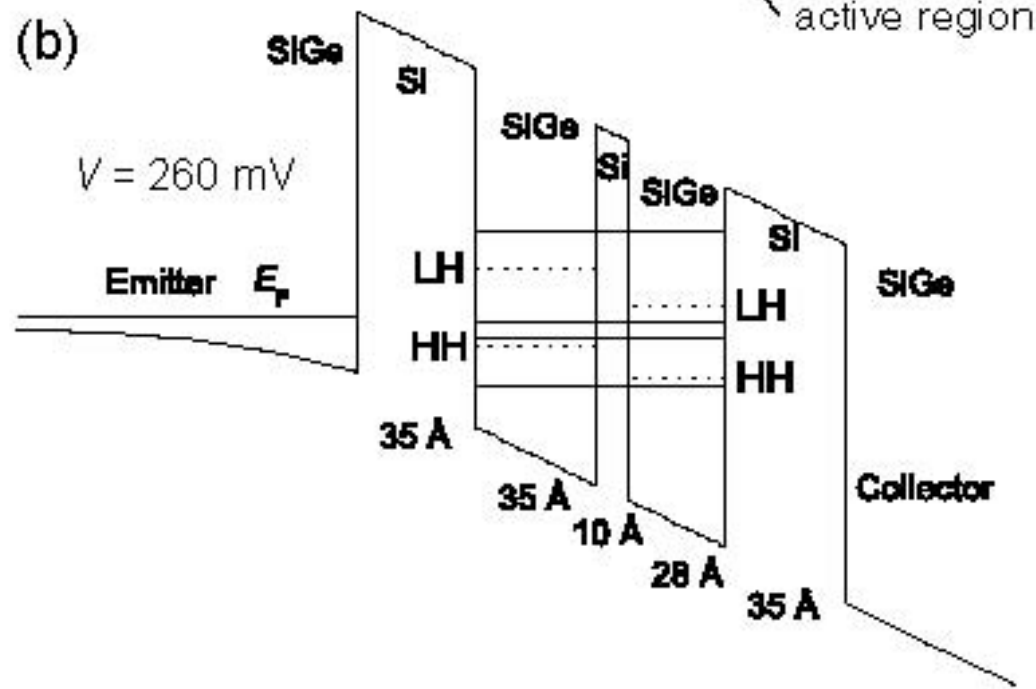

Fig. 1
Jun Liu et al.
Strain-induced quantum ring hole states in a gated vertical quantum dot

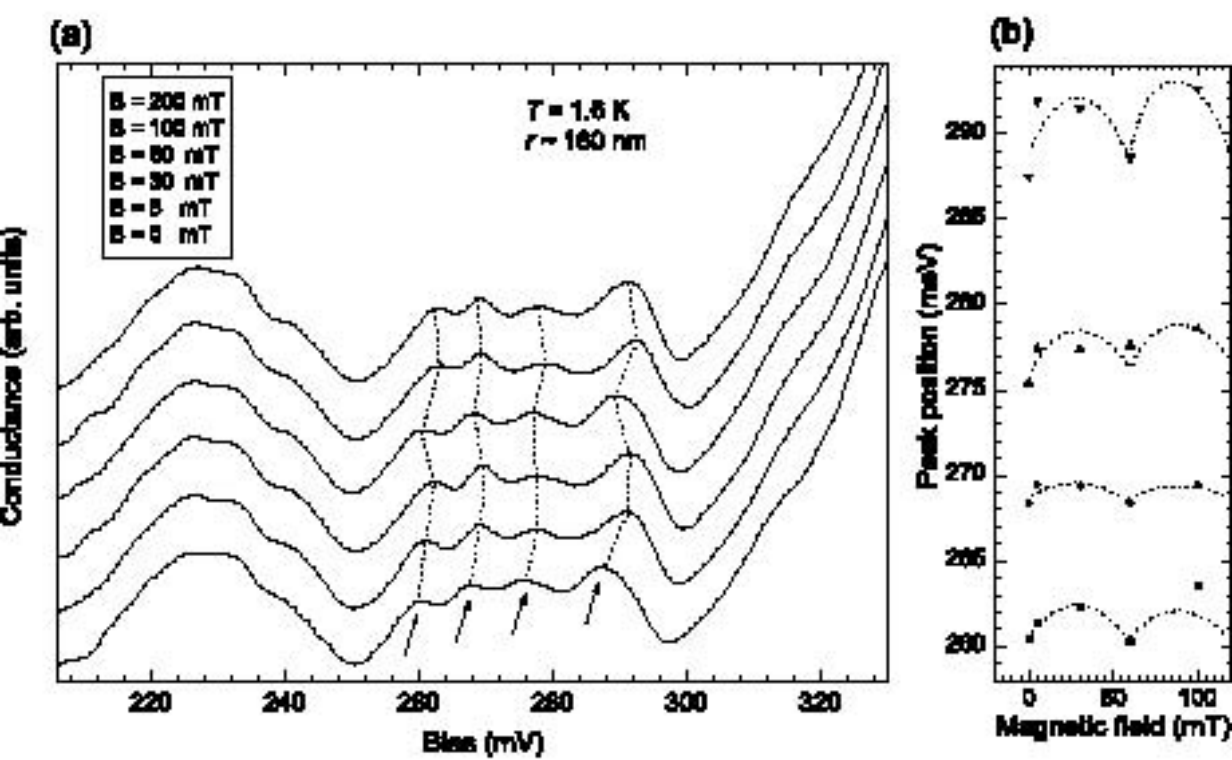

Fig. 2
Jun Liu *et al.*
Strain-induced quantum ring hole states in a gated vertical quantum dot

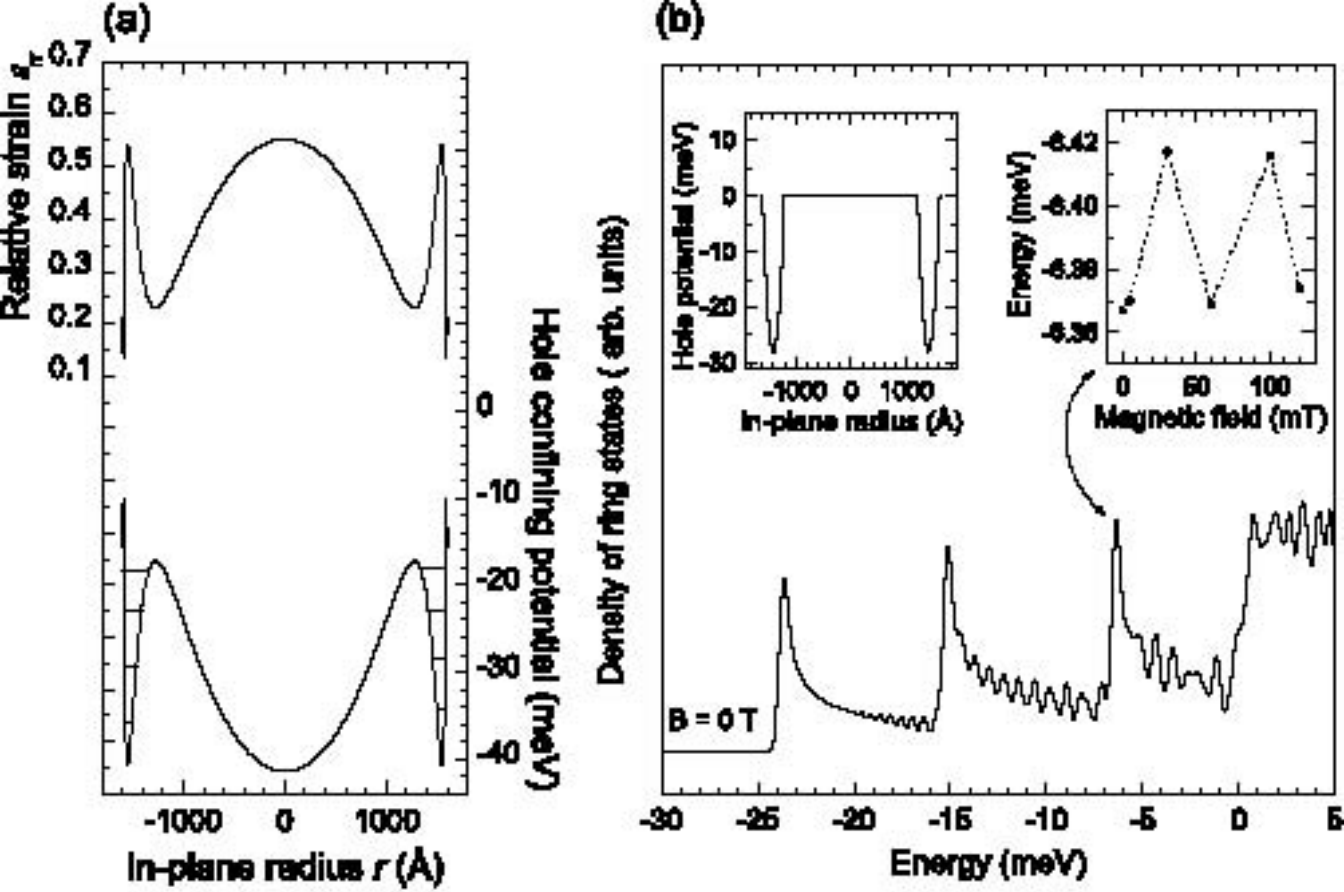

Fig. 3
Jun Liu et al.
Strain-induced quantum ring hole states in a gated vertical quantum dot

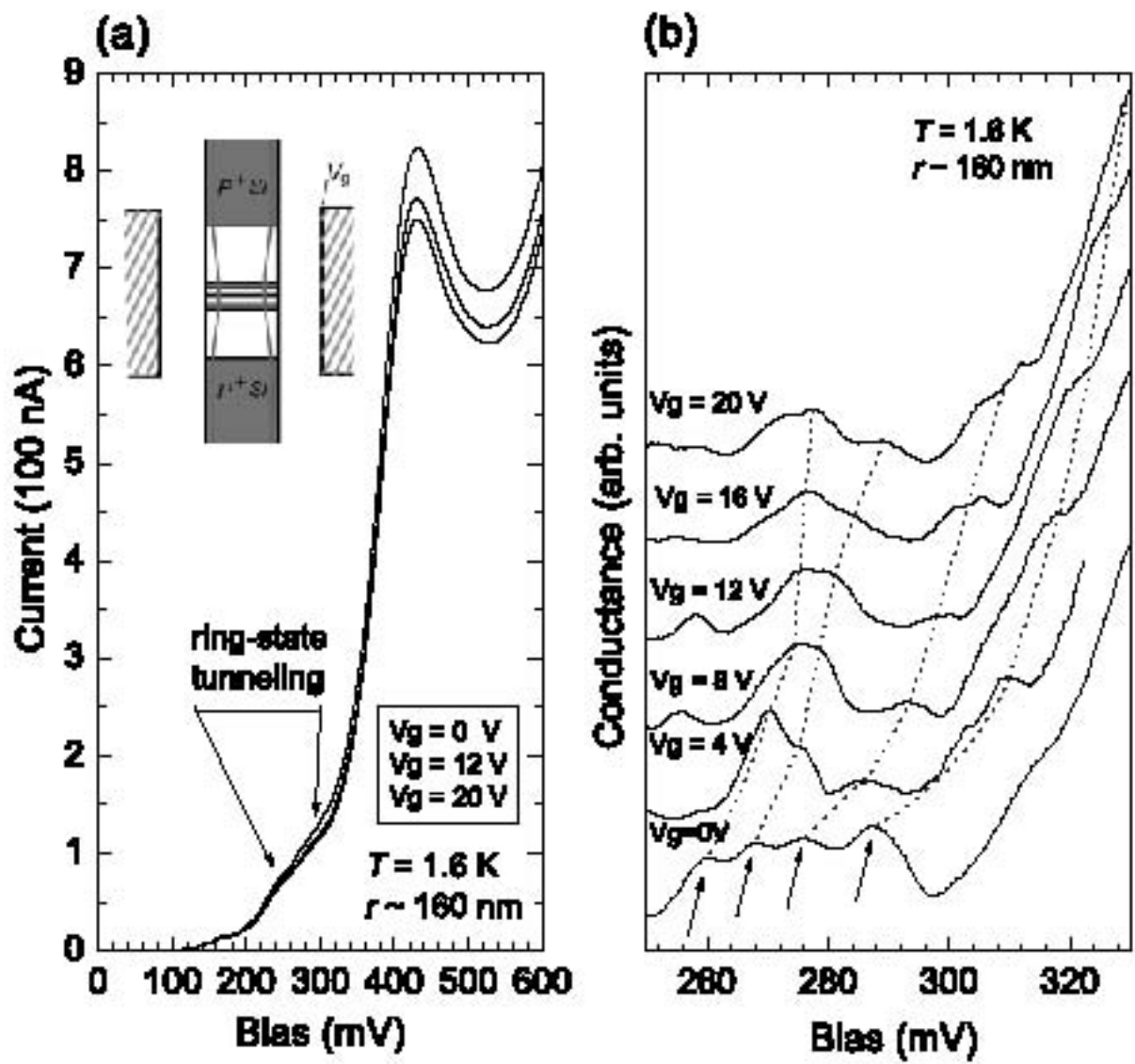

Fig. 4
Jun Liu et al.
Strain-induced quantum ring hole states in a gated vertical quantum dot